\documentclass[prd,aps,preprint,amsmath,nofootinbib,amssymb,eqsecnum,showkeys,tightenlines]{revtex4-1}
\usepackage{slashed}
\usepackage{epsfig,latexsym,cancel,amssymb,amsmath,verbatim,mathrsfs}
\usepackage{color}
\usepackage{graphicx}

\def\ra{\rightarrow}
\def\L{\left(}
\def\R{\right)}

\def\ld{\lambda}
\def\f{\frac}

\begin{document}

\title{WIMP Dark Matter Hidden behind its Companion}
\author{Jun Guo}
\email[E-mail: ]{jguo\_hep@163.com}
\affiliation{College of Physics and Communication Electronics, Jiangxi Normal University, Nanchang 330022, China}

\author{Zhaofeng Kang}
\email[E-mail: ]{zhaofengkang@gmail.com}
\affiliation{School of physics, Huazhong University of Science and Technology, Wuhan 430074, China}

\author{Peng Zhang}
\email[E-mail: ]{zhang\_peng961013@163.com}
\affiliation{School of physics, Huazhong University of Science and Technology, Wuhan 430074, China}

\date{\today}

\begin{abstract}

The WIMP dark matter (DM)  hypothesis now is in an awkward position, owing to the stronger and stronger exclusion from DM direct detection. In this article we design a mechanism to evade this  constraint. The idea is simple. DM has a companion, and they are both charged under the DM protecting symmetry $G$; they admit the trilinear coupling DM-DM-companion, so the  latter provides a portal to the standard model (SM) via, for instance, the coupling to Higgs doublet. Then, DM semi-annihilates into the companion to arrive correct relic density,  without leaving DM-nucleon scattering signal. The idea can be  realized for $Z_N$ symmetric models with $N>2$. We stress that this mechanism has the characteristics of co-annihilation, and as a matter of fact its effect becomes necessary near or above the TeV region. This means that it may be difficult to detect our dark matter directly or indirectly. 

\end{abstract}


\maketitle

\section{Introduction}

The existence of dark matter (DM) is one of the widely accepted evidences beyond the particle standard model (SM)~\cite{Bertone:2004pz}.  But so far, the only accurate quantitative information we have about DM is its  relic density, $\Omega h^2\simeq 0.12$~\cite{Planck:2018vyg}. This illustrative value inspires the well-known weakly interacting massive particle (WIMP) picture for DM, which is supposed to be a neutral particle around the weak scale, interacting with the SM members with a strength $\sim {\mathcal{O}}(g_2)$, thus demonstrating the WIMP miracle. In such a paradigm, barring parameters fine-tuning and peculiar structure for interactions, $\sigma_{\rm SI}$, the cross section of DM-nucleon spin-independent scattering should be large enough, leaving detectable signals which would have already been hunted by the underground experiments such as XENON~\cite{Aprile:2018dbl} and PandaX~\cite{Meng:2021mui}, etc. Nevertheless, we just see the experimental exclusion lines on $\sigma_{\rm SI}$ approaching the neutrino floor, below which DM signals will be buried in the neutrino backgrounds and become nearly undetectable. 

The decreasing upper bound on $\sigma_{\rm SI}$ means that we may have to face such a fact: the WIMP DM paradigm probably is not correct, at least does not assemble the typical one.  This may be a good news to the feebly interacting (with the  members of the SM) massive particle scenario~\cite{McDonald:2001vt,Hall:2009bx,Bernal:2017kxu,Kang:2014cia}, which gives up the freeze-out dynamics to determine DM relic density and assumes that DM has never reached thermal equilibrium with the SM bath, but it is still capable of producing correct relic density via the IR or UV-freeze-in mechanism~\cite{Chen:2017kvz}. Even within the WIMP framework, we still have many ways to maintain the WIMP miracle while avoiding the very  stringent direct detection bounds, for instance, the isospin-violating DM~\cite{Feng:2011vu,Kang:2010mh,Gao:2011ka}, and the pseudo-Goldstone DM~\cite{Gross:2017dan,Okada:2020zxo,Jiang:2019soj}. 

This article is devoted to exploring a new mechanism.  The idea is inspired by the structure of the Lagrangian for semi-annihilating DM models~\cite{DEramo:2010keq,Belanger:2012vp,Ko:2014nha}, to which a generalization admits the separation between DM annihilation cross section and $\sigma_{\rm SI}$. For illustration, let us consider the $Z_3$ symmetric model, where we introduce two identical scalars $S_{1,2}$, rather than merely one in the minimal version; both of them are transformed as $e^{ik2\pi/3}S_{1,2}$. Let $S_1$ be the DM, and its companion $S_2$ is properly heavier than $S_1$. These ingredients open a possibility to hide $S_1$ DM:  $S_1$ has a negligible Higgs portal coupling, so its scattering with nuclei is highly suppressed; the companion $S_2$ serves as the portal (dubbed as the companion portal) between $S_1$ and the SM Higgs sector, so the correct relic density of $S_1$ can be achieved by the semi-annihilation mode  $S_1S_1\rightarrow S_2(\ra S_1+\bar f f)h$. This structure can be easily generalized to $Z_N$ with $N>3$. In particular, if $S_1$ instead is a spin-1/2 fermion, its direct coupling with the SM Higgs sector is forbidden, naturalizing the  absence of $\sigma_{\rm SI}$, which is unlike the  previous version that requires one to squeeze the Higgs portal by hand. 
  
We must point out a related work~\cite{Saez:2021oxl}, which appeared on arxiv at the final stage of our article. Authors studied the assisted freeze-out mechanism~\cite{Belanger:2011ww}, which actually is a two components DM model with the dominant one $X_1$ feebly interacting with the SM while the subdominant one  $X_2$ opening the annihilation channel for $X_1$ freeze-out.  Their hypothesis has a similar picture to ours in treating the freeze-out dynamics of WIMP DM. However, the difference in the model building is obvious: we start from a single symmetry whereas they should impose more, and in our approach the companion for assisted freeze-out also transforms under this symmetry.  In fact, we find that the  assisted freeze-out mechanism can be easily integrated into our mechanism, for $N>3$. 

If direct detection is no longer feasible, we must rely on other detection methods, such as looking for relevant signals in cosmic rays or on colliders. However, this requires professional technology, which we do not fully have at present, so the detailed analysis is postponed to the follow-up work. For the indirect detection, the main difficulty is to obtain the precise spectra of DM annihilation, a pair of Higgs boson with one (on or off shell) from the companion cascade decay; such spectra is not regular and needs simulation. Note that our mechanism may have a strong co-annihilation effect, in particular for DM near the TeV scale, then the resulting WIMP DM is also hidden in the sky.


The paper is organized as follows: In Section II we present the mechanism and models. In Section III we   discuss the closely related phenomenology, focusing on the ways to obtain correct relic density, stressing the role of coannihilation. Section IV contains the conclusions and discussions. We add the details of calculations for the fermionic DM annihilation cross section in the appendix, which may involve some subtleties.

\section{The mechanism and the models}

Since our mechanism is motivated by the semi-annihilating DM models, we will start with the models with $Z_3$ and then generalize it to $Z_N$. 

\subsection{The $Z_3$ models}

\subsubsection{Model-A: Scalar DM with suppressed Higgs-portal by hand}

Let us demonstrate the mechanism based on the well-known $Z_3$ symmetric model, whose minimal form contains just a complex scalar $S_1$. But we will see one more such a scalar $S_2$ makes a real difference. For concreteness, both fields transform under $Z_3$ as $S_i\rightarrow e^{ik_i2\pi/3}S_i$ with $k_i=1$ or $2$, then this model  (model-A hereafter) has the (incomplete) Lagrangian incorporating the following terms
\begin{align}
 -{\cal L}_{Z_3}\supset &m_{1}^2S_1 S_1^* + m_{2}^2 S_2S_2^*\label{Z3L} + \lambda_{1h}|S_1|^2|H|^2+ \lambda_{2h}|S_2|^2|H|^2 
\\ \nonumber
&+\left( \frac{A_1 S_1^3}{3} +\frac{A_2 S_2^3}{3}+ \frac{1}{2}A_{12}S_1^2S_2 + \frac{1}{2}A_{21}S_1S_2^2 + c.c\right),\label{ModelA}
\end{align}
In the present work, we impose the CP symmetry, and therefore all couplings are real. Later, we will comment on the consequence of relaxing this assumption. $S_1$ and  $S_2$ receive mass terms both from the bare mass term $m_i^2 S_iS_i^*$ and the Higgs portal terms $\lambda_{ih}|S_i|^2|H|^2$, so their squared masses are given by
\begin{align}
m_{S_i}^2 = m_i^2 + \lambda_{ih}v^2,
\end{align}
where we have written $H=v+h/\sqrt{2}$ with $v=174$GeV. $S_1$ is assumed to be the lighter component thus the dark matter candidate, while $S_2$ can decay into $S_1+h(\rightarrow \bar f f)$ after including a term $\lambda_{12}S_1S_2^*|H|^2$, with the coupling strength $\lambda_{12}$ set to be sufficiently small, irrelevant to other dynamics~\footnote{As a matter of fact, $S_1$ and $S_2$ are indistinguishable and therefore in general they are expected to have a  mixing via the term $m_{12}^2S_1S_2^*$ and $\lambda_{12}S_1S_2^*|H|^2$; the mixing must be small, otherwise $S_1$ will have a sizeable coupling to the Higgs doublet, which should be suppressed in our mechanism. So, in writing the Lagrangian Eq.~(\ref{ModelA}), we are working in the mass basis, leaving the term $\lambda_{12}S_1S_2^*h$ for $S_2$ to decay.}.

The mechanism of hiding the WIMP dark matter candidate $S_1$ is simple. Let us assume that $S_1$ is properly heavier than its companion $S_2$, namely having mass within the window
\begin{align}\label{window}
m_{S_2}+m_h<2m_{S_1}<2m_{S_2}.
\end{align}
The window is wide, and it opens for the mass gap $\Delta=m_{S_2}-m_{S_1}$ satisfying  $0<\Delta<m_{S_1}-m_h$; DM mass must be heavier than the Higgs boson mass $m_h\approx 125$GeV. Eq.~(\ref{window})  ensures that the annihilation channel $S_1S_1\rightarrow S_2^* h$ opens, with a cross section proportional to $\propto A_{12}^2\lambda_{2h}^2$ rather than $\propto A_{12}^2\lambda_{1h}^2$. Thus, we can turn off the Higgs portal coupling of the dark matter candidate $S_1$ without turning off its annihilation channels. In this way, we disconnect the interactions for DM annihilating  from scattering with quarks, to avoid the extremely tight constraints from the latter. Note that the WIMP scenario is maintained, both $S_1$ and $S_2$ can be around the weak scale, with weak couplings strength. 

As a matter of fact, $S_2$ plays the role of mediator between the dark matter and the Higgs sector in the SM. But different than the usual models where the DM purely annihilate into a pair of mediator (may or may be not the Higgs boson pair), our model is characterized by the semi-mediator final states, due to the fact that the mediator is not neutral under the DM protecting symmetry.

\subsubsection{Model-B: Fermionic DM without Higgs portal}

One may be dissatisfied with the assumption that the Higgs portal coupling of $S_1$ should be very small. Is it possible to forbid it in a natural and simple way? To that end, we instead consider that $S_1$ is not a spin-0 but a spin-1/2 fermion, denoted as $\Psi_L$.  Its $Z_3$ charge is not changed. As a consequence, many unwanted terms go away due to the Lorentz symmetry. Owing to the $Z_3$ symmetry, $\Psi_L$ cannot be a Majorana fermion. So, to make it massive, the chiral partner $\Psi_R$ is needed.  Then, the general model takes the form of
\begin{align}\label{Z3:fermion}
 -{\cal L}_{Z_3}\supset & + m_S^2|S|^2 + M_\Psi\bar{\Psi}\Psi+ \lambda_{sh}|S|^2|H|^2\\
 \nonumber
 &+\L\frac{A_s}{3}S^3+\lambda_{L}\overline{\Psi^{C}}P_L\Psi{S}+\lambda_{R}\overline{\Psi^{C}}P_R\Psi{S}+c.c.\R,
\end{align}
where $\Psi=(\Psi_L,\Psi_R)$ is the Dirac fermion and the companion field $S_2$ is renamed as $S$; the projector $P_{L,R}=\frac{1\mp\gamma^5}{2}$ and $\Psi^{C}=C\Psi^T$ with $C=i\gamma^0\gamma^2$ is the charge conjugate field of $\Psi$. As expected, the direct connection between $\Psi$ and SM is absent, by virtue of the symmetry rather than by hand. 

This model (model-B hereafter) respects an accidental parity, namely the fermion number $\Psi\rightarrow-\Psi$. Therefore, both $S$ and $\psi$ are stable without introducing other interactions. But we have no interest in multi-component DM, and want to avoid such trouble. It is not a big problem, and in the proper extensions to the above minimal model, this accidental parity is supposed to be violated, admitting decay like $S\ra \Psi+\nu_L$. As proof of existence, let us consider the SM extended by the type-I seesaw mechanism with right handed neutrinos $\nu_R$, which are $Z_3$ neutral. Then couplings $S\bar\Psi_L \nu_R$ leads to the decay $S\ra \Psi_L+\nu_R$ if $\nu_R$ is sufficiently light, otherwise $S\ra \Psi_L+\nu_L$ via the mixing between $\nu_R$ and $\nu_L$~\footnote{It is of interest to consider a quite different realization of the mechanism within the type-I seesaw extension to SM. Now, the term like $\bar\Psi P_R \nu  S$ provide a portal to DM and opens the annihilation channel to a pair of sterile neutrinos. Such a way to avoid the direct detection bound heavily relies on the presence of $\nu_R$, and moreover even the $Z_2$ symmetry works. }. 

Another way to break the accidental symmetry is introducing an additional Higgs doublet $\eta$ with quantum number like $H$ but carrying one unit of $Z_3$ charge, and thus in general it mixes with $S$ due to the term $\eta^\dagger H S$; further, we couple it to the SM left-handed lepton doublet via $\eta\bar\Psi P_L L_i$, which then leads to the desired decays of $S$ and $\eta$ to $\Psi$. Although for different purposes, the Lagrangian Eq.~(\ref{Z3:fermion}), along with those terms to break the accidental symmetry for $\Psi$, is a subset of the model with radiative neutrino masses at two loop level~\cite{Ma:2007gq}. The impact of semi-annihilation is studied for the first time in Ref.~\cite{Aoki:2014cja}, which essentially covers our scenario of freeze-out  for $\Psi$. But their studies involve much more degrees of freedom and parameters, so, in this sense, our study is a simplification of their analysis. Another obvious difference is that, due to $\eta$ in the loop, the Dirac DM in their model gains effective vertex $\bar\Psi \sigma^{\mu\nu}\Psi F_{\mu\nu}$, which contributes to DM-nucleon scattering.

\subsection{Generalization to $Z_N$ or beyond}

From the previous model building, it is seen that the cubic term between DM and its companion fields, such as $S_1^2S_2$ and $\bar{\Psi}^{c}P_L\Psi{S}$, is the crucial structure for our mechanism to work, while the  portal term between the companion field and SM fields like$|S_2|^2|H|^2$ is trivial viewing from symmetry, and it can be replaced by other portals. Since those cubic terms are compatible with $Z_3$, it explains why we start from this symmetry. But the cubic term is allowed by many other symmetries, and this subsection is devoted to exploring simple generalizations of $Z_3$.

It is illustrative to analyze the model from the pattern of symmetry, which is helpful to generalize the mechanism from $Z_3$ to other possible symmetry $G$. Here, the annihilation of DM is characterized by non-$G$ invariant: The initiate state is the DM-DM pair, carrying a non-vanishing $G-$charge, and the final states, containing the companion field, are not $G$ invariant neither, carrying the same $G-$charge. Such a situation is in contrast to the familiar case like $Z_2$, where the initial DM pair and the final states must be $Z_2$ invariant. Although the minimal, $Z_3$ is definitely not the unique symmetry to realize our scenario of DM annihilation. For simplicity, let us focus on the case that $G$ is an Abelian symmetry $Z_N={e^{i2\pi k/N}} (k=0,1,...N-1)$with $N>3$. In the following, we will derive the condition for $Z_N$ to work. Suppose the DM and its companion field carry $Z_N$ charges $k_1$ and $k_2$, respectively. Therefore, charge conservation yields the following equation 
\begin{align}\label{ZN:assign}
2k_1+k_2=0~{\rm module} ~N, ~~(k_i=1,...N-1).
\end{align}
In the language of Lagrangian, one readily sees that the crucial cubic term $S_1^2S_2$ along with the trivial Higgs portal term $|S_2|^2|H|^2$ are always present (Obviously, it is true also for the fermionic model.). For $N > 3$, the equation admits many solutions, for instance, 
\begin{align} 
&N=4:~~k_1=1,~k_2=2;  \quad k_1=3,~k_2=2,\\
&N=5:~~k_1=1,~k_2=3; \quad k_1=2,~k_2=1; \quad k_1=3,~k_2=4; \quad k_1=4,~k_2=2;\\
&N=6:~~k_1=1,~k_2=4; \quad k_1=2,~k_2=2; \quad k_1=4,~k_2=4; \quad k_1=5,~k_2=2.
\end{align}
By the way, for $N=3$, the solutions are $k_1=k_2=1$ or $2$, thus in either case the two fields carrying the same charge~\footnote{One can prove that only the $Z_N$ which has the $Z_3$ subgroup admits such solutions: Let the solution be $1\leq k_1=k_2=k=nN/3\leq N-1$, with $n=1,2$. So, to make $k$ be an integer, one must require $N=3m$. Then, in $Z_{3m}$, the two solutions $k=m$ and $2m$ exactly correspond to the two solutions in $Z_3$.}.

But generically speaking, for $N>3$, these solutions require two fields with different charges. As a consequence, they can be distinguished and cannot be mixed with each other even if the spins are the same (both are scalar fields). Then it is likely that both particles will be stable, leading to multi-component dark matter; a related study is Ref.~\cite{Yaguna:2019cvp}. For concreteness, let us show a model for $N=5$, with two scalars $S_1$ and $S_2$ which respectively carry charges $k_1=1$ and $k_2=3$. Then, the interactions of the most general (also typical for the larger $N$) renormalizable Lagrangian is
\begin{align} 
{\cal L}_{Z^5}=\lambda_1|S_1|^2|H|^2+\lambda_2|S_2|^2|H|^2+\L A_{12}S_2^2S_1+c.c.\R,
\end{align}
$S_2$ can decay only via the term $S_1^2S_2$, however, it is kinematically forbidden in the mass window Eq.~(\ref{window}) we are considering. In this case, we must consider the rather strong coupling $\ld_2$, so that the annihilation cross section of the companion $S_2$ is large enough to make its final abundance sufficiently small. Interestingly, this picture is nothing but just the one mentioned in the introduction~\cite{Saez:2021oxl,Belanger:2011ww}.

Note for the higher $N$ case, some irrelevant terms such as $S_i^3$ may be forbidden by symmetry, but other irrelevant terms like $S^4_i$ may arise. Anyway, in addition to the crucial terms, the other terms appearing in the Lagrangian depends on the concrete $N$ and as well charge assignments.

Any $Z_N$ discrete symmetry can be a residual symmetry of $U(1)$; see an example~\cite{Guo:2015lxa}. Therefore, $G$ can be $U(1)$, either global or local. In particular, if $U(1)$ is not spontaneously broken, then Eq.~(\ref{ZN:assign}) should be understood as the charge neutrality condition.

\section{Phenomenology of the DM-companion dark sector}


In our mechanism, the main concern on DM is its relic density. In the present study, we focus on the $Z_3$-like case, where the companion can decay. Short comments on the possible probes to the model are made.  
 
\subsection{Freeze-out via semi-annihilation, including co-annihilation}

\subsubsection{Not standard semi-annihilation}

In our models, the annihilation modes for DM are very limited, and in either model, they are dominated by a pair of DM particles annihilating into the SM Higgs boson plus the companion particle. Since the companion will eventually contribute to the final DM relic density, essentially the DM freezes out by means of semi-annihilation, which halves the efficiency of DM annihilation. 

In the model-A, the Feymann diagrams for DM semi-annihilation are shown in Fig.~\ref{indirect:s}. There are two modes. We did not show the ordinary semi-annihilation mode, $S_1 S_1 \rightarrow h S_1^*$, which is supposed to be negligible because of the suppressed Higgs portal for DM. But in the numerical study we will also include this contribution, to see if the state-of-the-art DM direct detection still allows a pure Higgs portal DM scenario. The left panel shows the dominant mode,  $S_1 S_1 \rightarrow h S_2^*$ in the $s$-channel exchange of the companion. Its cross section $\sigma_{s,1}$ times the DM relative velocity is give by
\begin{align}\label{sigma:s}
\sigma_{s,1} v_r = \frac{1}{64\pi}\frac{|\vec{p}_h|}{m_{S_1}^3}\frac{|A_{12}|^2 (\lambda_{2h} v)^2}{(4m_{S_1}^2 - m_{S_2}^2)^2},
\end{align}
where ${\vec{p}}_h$ is the momentum of the Higgs boson, in the center-of-mass (CM) frame, taking the form of
\begin{align}\label{ph}
|\vec{p}_h|
\approx m_{S_1}\ld(1,m_{S_2}^2/4m_{S_1}^2,m_{h}^2/4m_{S_1}^2),
\end{align}
with $\ld(1,x,y)=\sqrt{(1-x-y)^2-4xy}$, which usually results in a moderate phase space suppression. One can rewrite Eq.~(\ref{sigma:s}) as 
\begin{align}\label{sigma1:s}
\sigma_{s,1} v_r = \frac{1}{64\pi}\f{|A_{12}|^2 (\lambda_{2h} v)^2}{16m_{S_1}^6}f(1,m_{S_2}^2/4m_{S_1}^2,m_{h}^2/4m_{S_1}^2),
\end{align}
where $f(x,y)=\ld(1,x,y)/(1-x)^2$ is an order one number except for $x\ra 1$, hitting the resonant pole. Since we are considering CP conservation in the dark sector, the CP conjugate process $S^*_1 S^*_1 \rightarrow h S_2$ has an equal annihilation rate. At the end of this article, we will briefly discuss the consequence if this is not true.
\begin{figure}[htbp]
\centering
\includegraphics[width=0.4\textwidth]{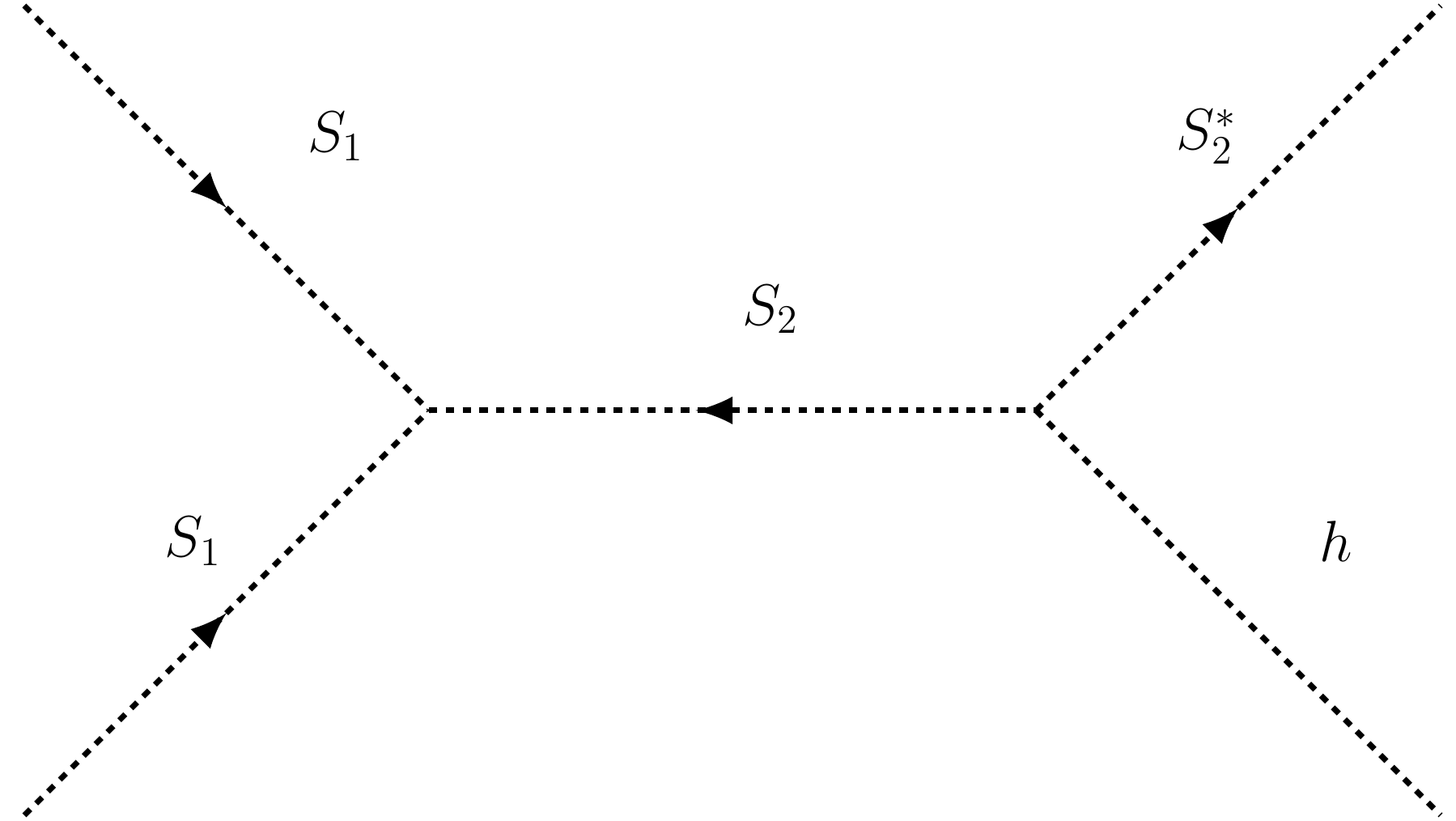}\quad\quad\quad
\includegraphics[width=0.38\textwidth]{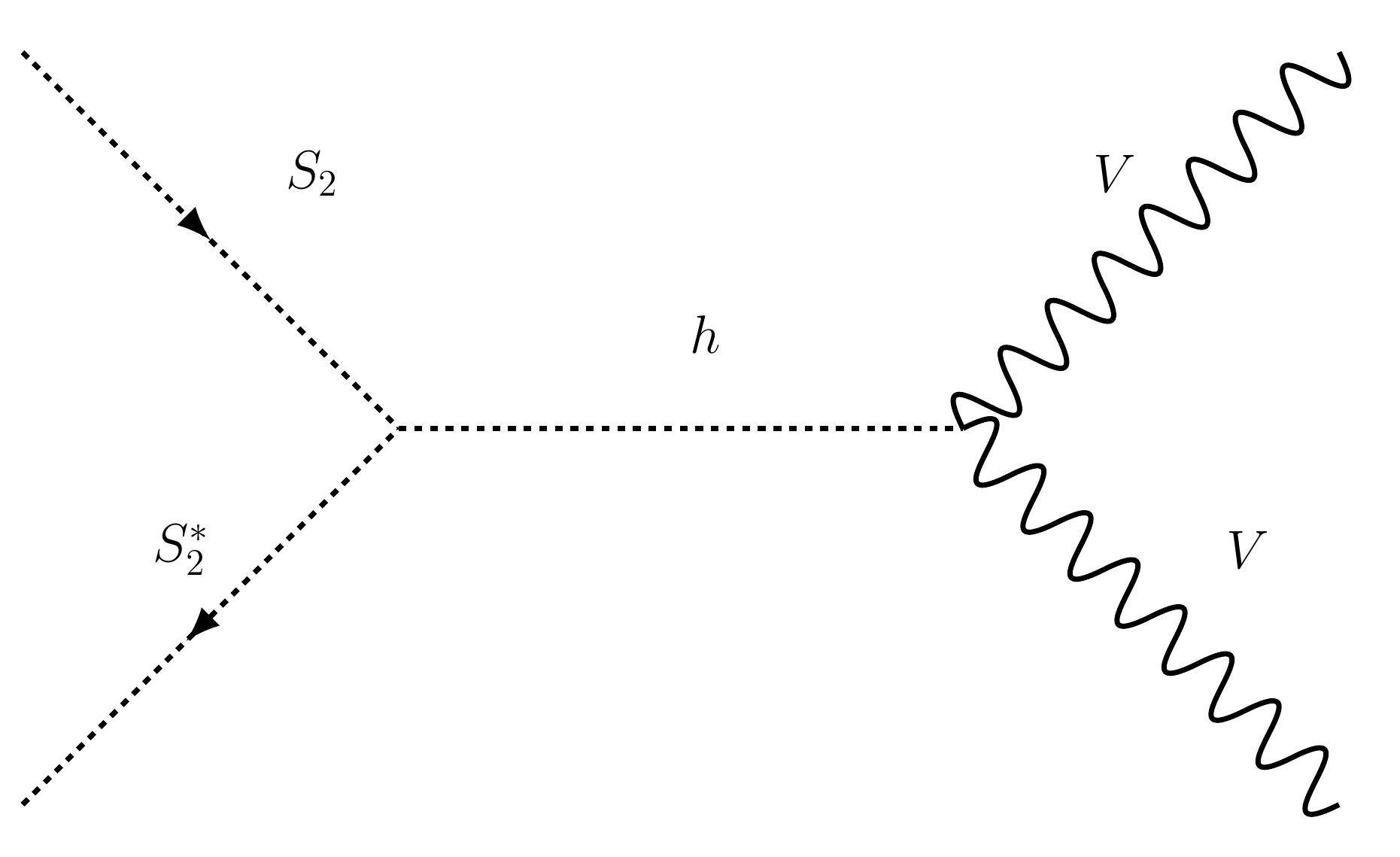}
\caption{Left:  The Feymann diagram for DM semi-annihilation $S_1S_1\ra S_2^*h$ in the $s$-channel and DM-companion coannihilation  $S_1S_2^*\ra S_1^*h$ in the $t$-channel; Right: The diagram for companion annihilation into the SM memebers such as the vector bosons via the Higgs-portal.}
\label{indirect:s}
\end{figure}

In the model-B,  the DM is instead a Dirac fermion $\Psi$, its only annihilation mode is $\Psi\Psi \rightarrow hS^* $, in the $s$-channel exchange of the companion. In the appendix.~\ref{cross:calu}, we give the details of calculating the annihilation cross section $\sigma_{f,1}$ times the DM relative velocity $v_r$:
\begin{align}\label{sigma:f}
\sigma_{f,1} v_r 
=\f{1}{64\pi }\f{(\lambda_L-\lambda_R)^2(\lambda_{sh}v)^2}{16M_\Psi^4}
f(1,m_{S}^2/4M_\Psi^2,m_{h}^2/4M_\Psi^2).
\end{align}
This cross section is almost identical to the one given in Eq.~(\ref{sigma:s}), except that here the parameter $(\ld_L-\ld_R)M_\Psi$  plays the role of $A_{12}$.  In addition, $\sigma_{f,1} v_r$ anishes as $\ld_L\ra\ld_R$, due to the interference effect. One can understand this limit by combining the  two Yukawa couplings in the Lagrangian Eq.~(\ref{Z3:fermion}), to yield a Majorana-like Yukawa coupling $\ld \overline{\Psi^{C}}\Psi{S}+c.c.$; such a coupling is well-known to give an annihilation cross section without $s$-wave. 

\subsubsection{Impacts of DM-companion co-annihilation}

Although semi-annihilation has been widely studied, in our mechanism, it is additionally characterized by the possibility of co-annihilation~\cite{Griest:1990kh} due to the presence of companion. Despite of the more or less tuning of parameters, the physical consequence of semi \& co-annihilation DM is still very significant. The co-annihilation modes involving the companion in the initial state are inactive today, and therefore, if co-annihilation is the dominant  mechanism to determine DM relic density, the DM detectors, not only direct detectors installed in the ground, but also indirect detectors flying in the sky, turn a blind eye to DM. 
 
 In the following, we analyze the underlying reasons for the presence of co-annihilation, and as well its impacts to the ordinary semi-annihilation. Our example is again model-A, but the discussions completely apply to model-B. 
\begin{itemize}
\item  First, the mass of the companion is supposed to be close to that of the DM, in particular for DM mass not far above the Higgs boson mass, which is clearly seen from the allowed window: $\Delta/m_{S_1}<1-m_h/m_{S_1}$. The coannihilation effect becomes  nonnegligible when the ratio $\Delta/m_{S_1}\lesssim 1/x_f\sim 5\%-10\%$, with $x_f=m_{S_1}/T_f\sim 20$ denoting the decoupling temperature of DM freeze-out; the freeze-out process of two almost degenerate thermal particles  cannot be simply separated. 
\end{itemize}
\begin{itemize}
\item Next, the DM-companion co-annihilation mode is simply from the Feymann diagram that leads to DM-DM annihilation; see the left panel of Fig.~\ref{indirect:s}. The  co-annihilation $S_1S_2\ra S^*_1 h$ is just the $t$-channel exchange of the companion. We denote its cross section as $\sigma_{s,2}$, which is related to  $\sigma_{s,1}$ by the cross symmetry, and therefor they have almost equal area. This means that its impact is not very dramatic, considering the exponential suppressing factor shown in Eq.~(\ref{sigmaeff}). 
\end{itemize}

\begin{itemize}
\item Moreover, $\sigma_{3}$, the cross section of the companion-companion annihilation $S_2S_2^*\ra hh/VV/f\bar f$ with $V=Z$ or $W$ (see the right panel of Fig.~\ref{indirect:s}), can be enhanced by its relatively strong Higgs portal coupling and the  trilinear term coupling $A_2$~\footnote{Such a strong Higgs portal is beneficial to make the electroweak phase transition strong first order, a sufficient condition for the electroweak baryogenesis mechanism. As a matter of fact, in the  $Z_3$ symmetric model, this possibility has been studied by several groups~\cite{Kang:2017mkl,Kannike:2019mzk,Chiang:2019oms}. In addition to that, if the companion is also the component of DM relic, which probably is the situation in the models with $Z_N$ with $N>3$, the large coupling is required to sufficiently reduce the fraction of companion in the total DM relic density.\label{EWSB}}, and consequently it may even dominate the effective cross section. Note that the companion Higgs portal thus $\sigma_3$ is common to model-A and -B, so we do not need a further subscript to distinguish the two models. 
\end{itemize}

The precise relic density of DM is determined by the Boltzmann equations which describe the evolution of DM number density. In the coannihilation scenario, the relevant number density is the sum of DM and companion, $n=n_{S_1}+n_{S_2}$, and its evolution resembles the ordinary one, with the cross section replaced by the effective cross section,
\begin{align}
\frac{\mathrm{d}n}{\mathrm{d}t}&=-3H n-\langle\sigma_{eff}v\rangle(n^2 - n_{eq}^2),\\
\sigma_{eff}&\simeq \frac{1}{2}\sigma_{s,1}+\sigma_{s,2}(1+\delta)^{\frac{3}{2}}e^{-x_f\delta}+\sigma_3(1+\delta)e^{-2x_f\delta},\label{sigmaeff}
\end{align}
where $H$ is the Hubble parameter. The cross sections $\sigma_{s,1/2}$ and $\sigma_3$ are  introduced previously, and in  $\sigma_{eff}$ the contributions from $\sigma_{s,1/2}$ have been halved, because each semi-annihilation just reduces one dark particle (DM or the companion). The ratio $\delta\equiv\Delta/m_{S_1}\ll 1$ measures the DM-companion mass degeneracy. If their mass difference is large, the co-annihilation effect is not important and then the Boltzmann equation above will reduce to the standard form.




\subsection{Comments on probes to the dark sector}

Although in the model-B the fermionic DM candidate $\Psi$ does not scatter with the atom, in the model-A, the scalar DM candidate $S_1$ does, due to the non-vanishing Higgs portal coupling $\lambda_{1h}|S_1|^2|H|^2$. Then, the Higgs boson mediates $S_1$-nucleon spin-independent scattering, with a cross section estimated to be~\cite{Jungman:1995df}
\begin{align}\label{SI:H}
\sigma^{(n)}_{\rm S_1}=\frac{\lambda_{1h}^2}{4\pi}\frac{ \mu_{n}^2 m_n^2}{m_{S_1}^2m_h^4}\left(\sum_{q=u,d,s} f^{(n)}_{T_q}+\frac{2}{9}f^{(n)}_{T_G}\right)^2,
\end{align}
where $\mu_{n}\simeq m_n$ is the DM-nucleon reduced mass, and $f_{T_q}^{(n)}$ the nucleon parameters. In the case that the companion is also a tiny component of dark matter, the companion-nucleon scattering has a cross section similar to the above expression. But the corresponding contribution to the event rate should be multiplied by the fraction of this component.  

In general, our dark matter candidate leaves hints in the cosmic ray via the following signature: 
\begin{align}
    {\rm DM}+{\rm DM}\ra {\rm companion}(\ra {\rm DM}+h~~ {\rm or}~~ h^*)+h,
\end{align}
where the Higgs boson from the companion decay may be on-shell or off shell, depending on the mass splitting $\Delta $. In either case, the resulting cosmic ray spectra, such as of anti-proton and photon, need detailed simulations, which is beyond the current project and will be present elsewhere. As a fairly rough approximation, we may use the spectra from $ {\rm DM}+{\rm DM}\ra h+h$, and there are available constraints in the literature~\cite{Abazajian:2020tww}. Later, we will show this schematic constraint. Note that the indirect detection signal of DM may be greatly weakened in the co-annihilation mentioned region.

If the companion has a relatively strong interaction with the Higgs sector, the dark sector may leave imprints at the collider such as LHC. As noted in footnote~\ref{EWSB}, such a companion can play the role of trigger for strongly first order of electroweak phase transition. Then, one can probe its effects on the Higgs physics by measuring the Higgs self-coupling, which has been widely discussed. But this approach can not distinguish different bosonic triggers. To identify the companion, it is better to produce it directly at the collider. They can be pairly produced via gluon gluon fusion at the LHC, giving rise to signals of missing energy plus a pair of (on shell or off shell) Higgs bosons. Note that similar signals are present in supersymmetry, where the next-to-lightest sparticle (LSP) may decay into the LSP plus Higgs boson~\cite{Martin:1997ns}. However, if the companion itself behaves as missing energy at the detector, one should turn to different strategies.

\subsection{Parameter Scan and Result}

We display the numerical results of the two benchmark models, using the program micrOMEGAs~\cite{Belanger:2020gnr}. The model-A given in  Eq.~(\ref{ModelA}) contains 8 parameters, but only four of them are relevant, which are chosen to be: 1) masses of DM and the companion, $m_i = \sqrt{M_i^2 + \lambda_{ih}v^2}$; 2)  the companion-Higgs portal coupling $\lambda_{2h}\sim{\cal O}(1)$, whereas the DM-Higgs portal coupling is suppressed $\lambda_{1h}\ll 1$; 3) the DM-companion coupling $A_{12}$ and the trilinear term coupling for the companion,  $A_2$, which is relevant  in the coannihilation scenario to enhance the annihilation cross section for the companion. Similar analysis on the parameter space apply to model-B, and one can find that it also involves four key parameters. To learn how well the parameter space of our model fits the experimental observations, we use Metropolis-Hastings algorithm~\cite{mickay} to scan the parameter space. In Table.~\ref{pspace:s} and Table.~\ref{pspace:f} we show the range and the random walk size for each of parameters. 
~\\

\begin{minipage}{\textwidth}
 \begin{minipage}[t]{0.45\textwidth}
  \centering
       \begin{tabular}{|c|c|c|} 
\hline \multicolumn{3}{|c|}{model-A}\\
  \hline
  parameters & range & step size \\
  \hline
   $A_1$ & $10 \rm\ GeV$ & $0 \rm\ GeV$\\
  \hline
   $A_2$ & $[0, 1000]\rm\ GeV$ & $200 \rm\ GeV$\\
  \hline
   $A_{12}$ & $[0, 1000]\ \rm GeV$ & $200 \rm\ GeV$\\
  \hline
   $A_{21}$ & $10\rm\ GeV$ & $0 \rm\ GeV$\\
  \hline
   $\log_{10}\lambda_{1h}$ & $[-3, 0]$ & $1$\\
  \hline
   $\lambda_{2h}$ & $[0.1, 2.5]$ & $0.6$\\
  \hline
   $m_{S_1}$ & $[m_h, 2000]\rm\ GeV$ & $400 \rm\ GeV$\\
  \hline
   $\log_{10} \delta$ & $[-2, 0]$ & $1 $\\
  \hline
 \end{tabular}
\makeatletter\def\@captype{table}\makeatother\caption{Information of the scanned parameters with  $\delta = \frac{m_{S_2}}{m_{S_1}}-1$}\label{pspace:s}
 \quad\quad\quad\quad\quad\quad
  \end{minipage}
  \begin{minipage}[t]{0.45\textwidth}
   \centering
        
         \begin{tabular}{|c|c|c|}        
         \hline \multicolumn{3}{|c|}{model-B}\\
  \hline
  parameters & range & step size \\
 
  \hline
   $A_S$ & $[0, 1000]\rm\ GeV$ & $200 \rm\ GeV$\\
 
  \hline
   $\lambda_{sh}$ & $[0.1, 2.5]$ & $0.6$\\
  \hline
   $\lambda_{L}$ & $[0.1, 2.5]$ & $0.6$\\
  \hline
   $\lambda_{R}$ & $0.1$ & $0$\\

  \hline
   $M_{\Psi}$ & $[m_h, 2000]\rm\ GeV$ & $400 \rm\ GeV$\\
  \hline
   $\log_{10} \delta$ & $[-2, 0]$ & $1 $\\
\hline
   $  $ & $ $ & $ $\\
\hline
   $ $ & $ $ & $  $\\
  \hline
      \end{tabular}
       \makeatletter\def\@captype{table}\makeatother\caption{Information of the scanned parameters with $\delta = \frac{m_{S}}{m_{\Psi}}-1$ }\label{pspace:f}
   \end{minipage}
\end{minipage}

The experimental observations we consider are relic density  and two most stringent DM direct detection experimental results, XENON.~\cite{Aprile:2018dbl} and PandaX.~\cite{Meng:2021mui}.  We set the observed relic density in log scale $\log_{10}\widehat{\Omega h^2} = -1$; to get a larger allowed parameter space, the uncertainty of $\log_{10}\Omega h^2$ is set to be $\sigma = 0.5$. Then, the posterior log-likelihood for a certain parameter point could be written as:
\begin{align}
\ln P(Obs|x) = \ln\frac{1}{\sqrt{2\pi}\sigma} \exp\left(\frac{(\log_{10}\widehat{\Omega h^2} - \log_{10}\Omega h^2)^2}{2\sigma^2}\right) + \ln(p_{val}),
\label{loglikelihood}
\end{align}
where $p_{val}$ is the probability of acceptance by all recent DM direct detection experiments, and it is
calculated by  micrOMEGAs~\cite{Belanger:2020gnr} using the maxsimum gap method~\cite{gap method}. After applying MCMC process for $1\times 10^6$ times, we get the posterior distribution of parameter 
space, and we select the last $1\times 10^5$ points for analysis.

\begin{figure}[htbp]
\centering
\includegraphics[width=0.4\textwidth]{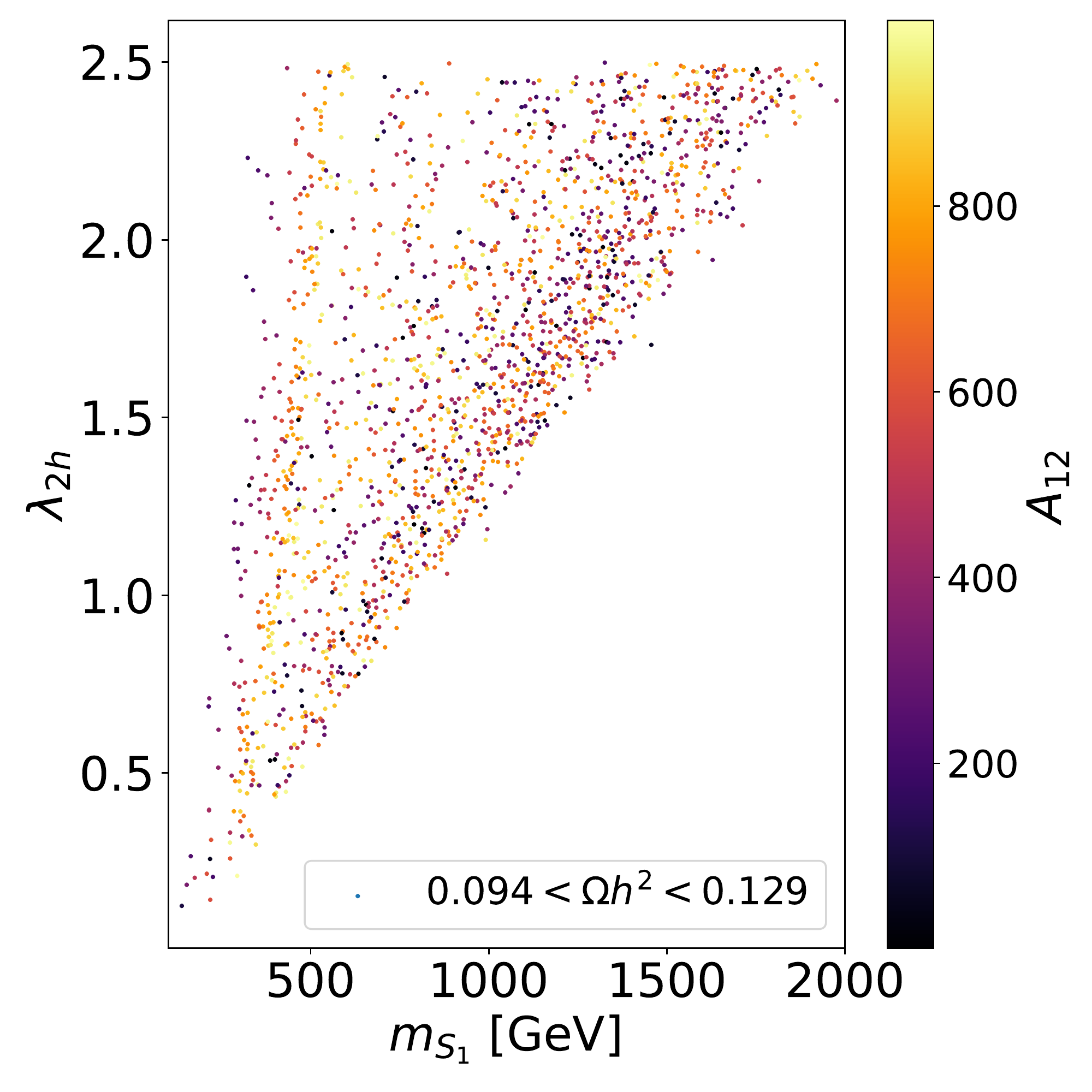}
\includegraphics[width=0.4\textwidth]{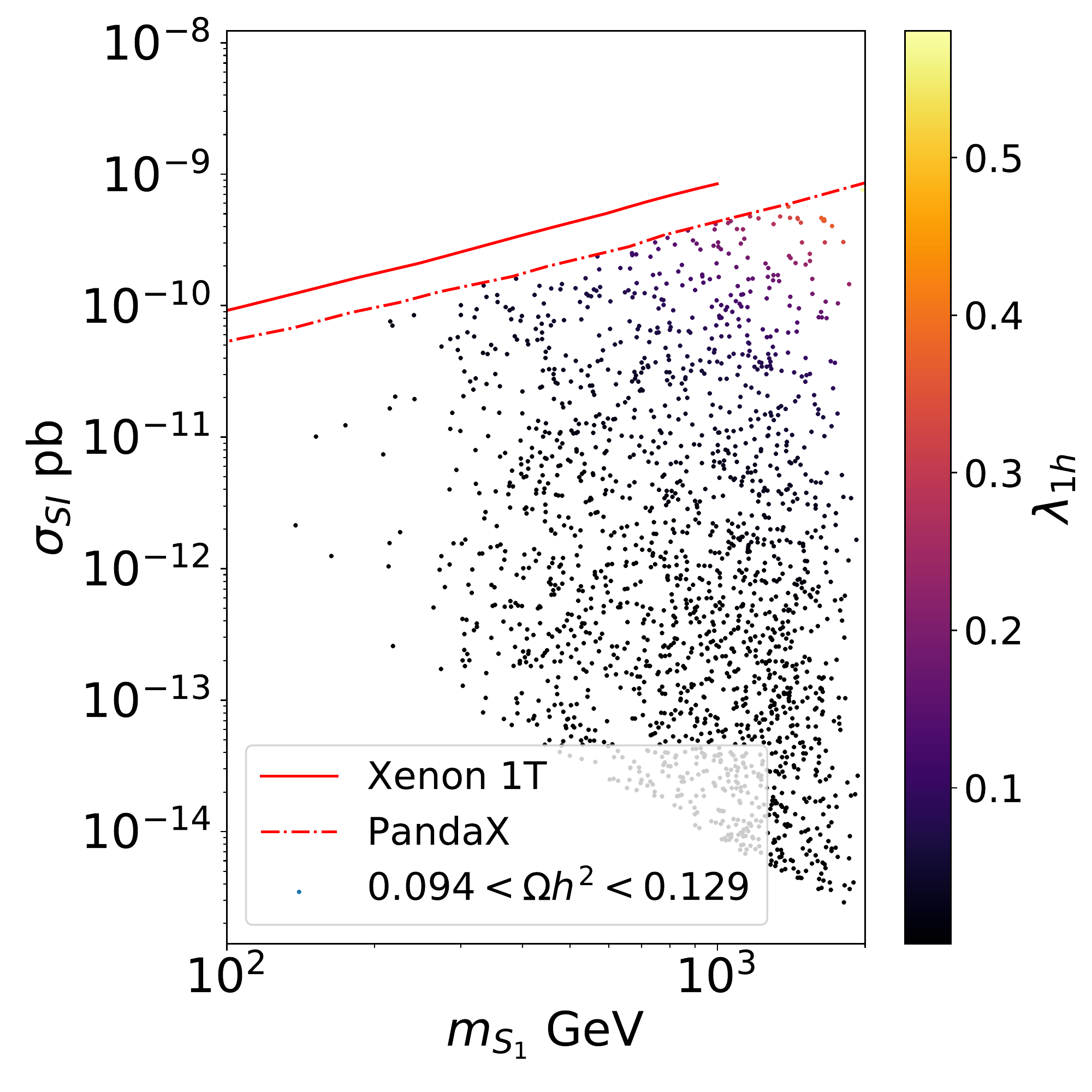}
\caption{Correct DM relic density in model-A. Left: the viable parameter space on the $m_{S_1}-\ld_{2h}$ plane, with colors encoding the value of DM-companion coupling $A_{12}$. Right: for the low value of $\lambda_{1h}$, 
our model escape the strong constrain of Xenon 1T
 }
\label{para:s}
\end{figure}

\begin{figure}[htbp]
\centering
\includegraphics[width=0.4\textwidth]{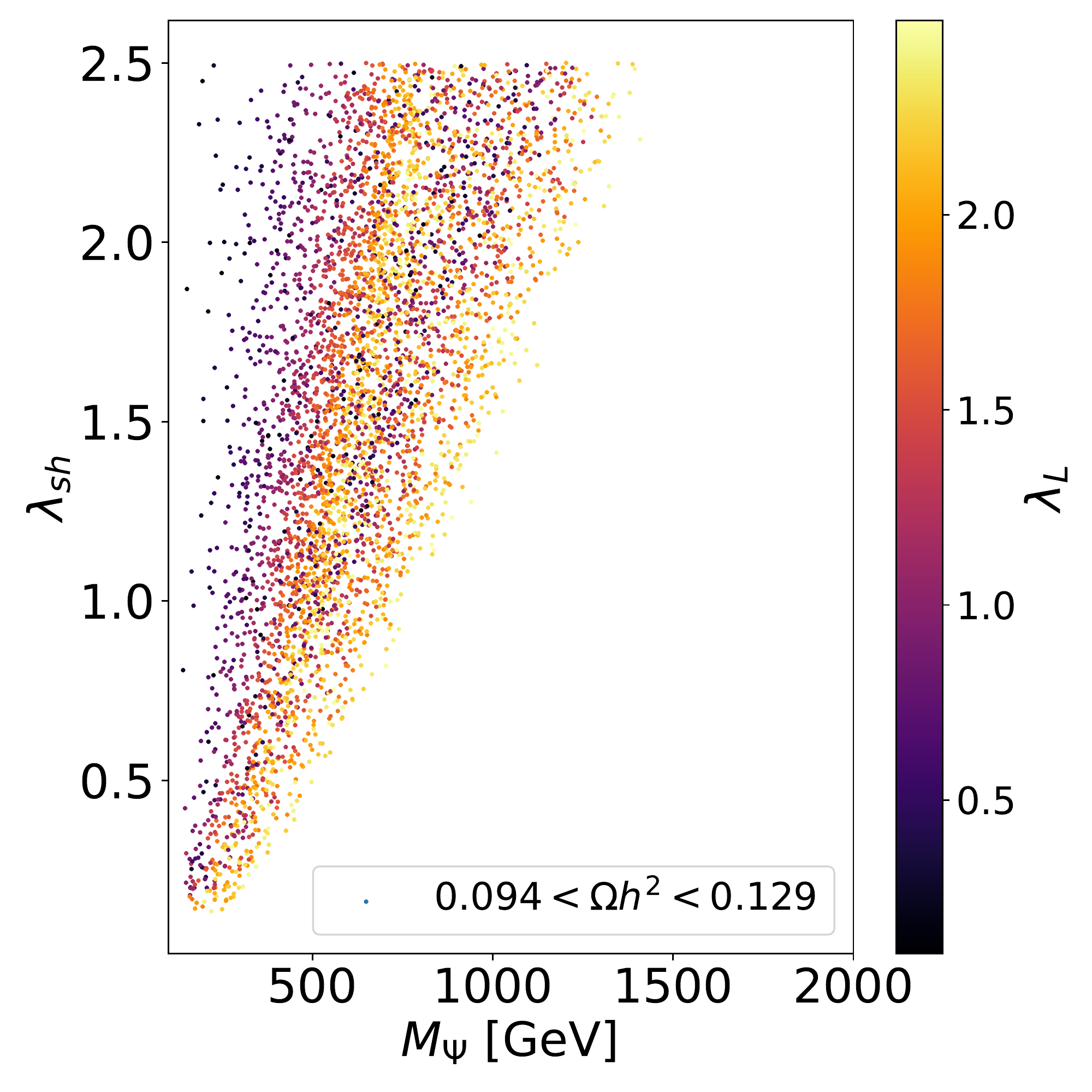}
\includegraphics[width=0.4\textwidth]{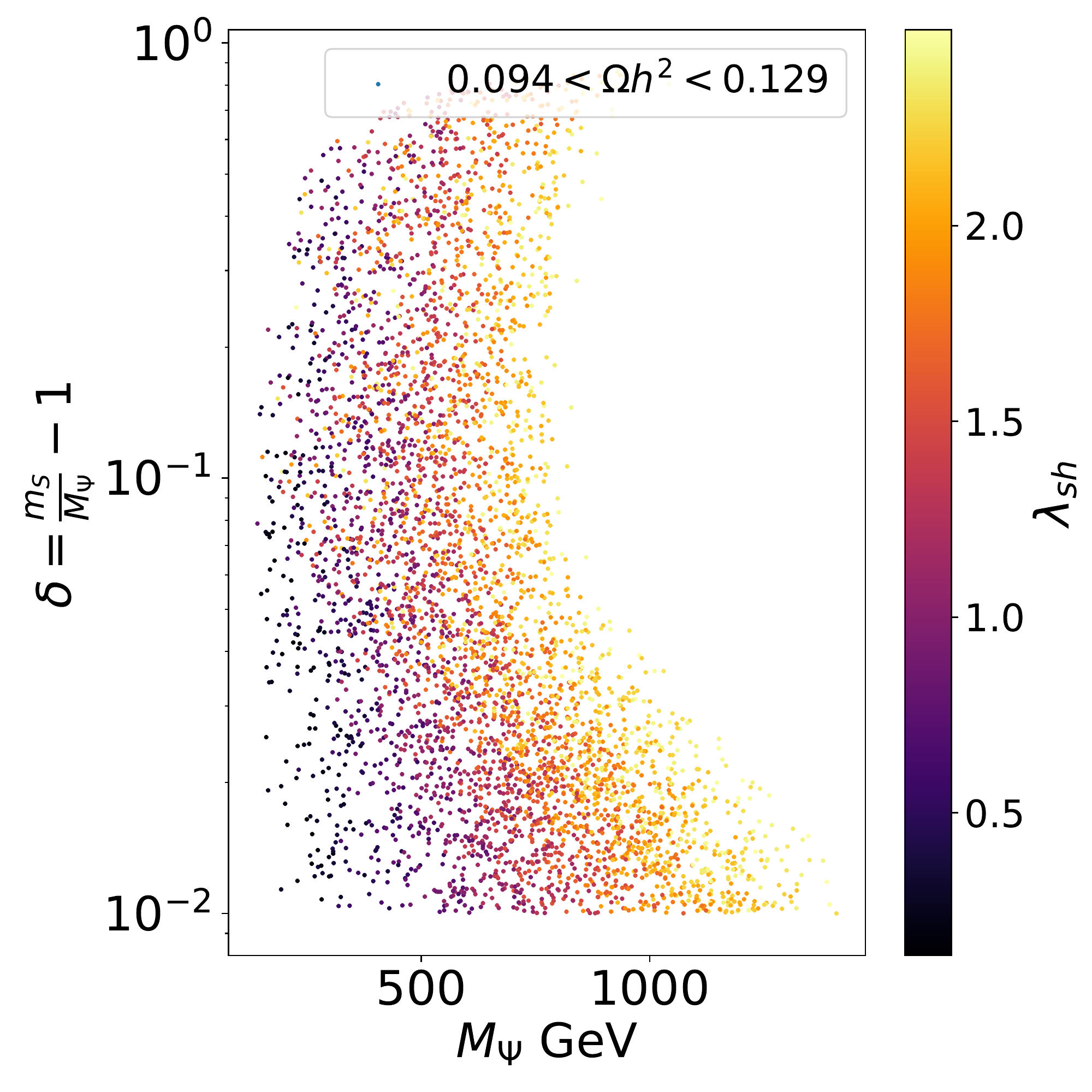}
\caption{Correct DM relic density in model-B. Left: the viable parameter space on the $M_\Psi-\ld_{sh}$ plane, with colors encoding the value of $\ld_L$. Right: Plots on the $M_\Psi-\delta$ plane with colors encoding the value of $\ld_{sh}$.
 }
\label{para:f}
\end{figure}

Let us end up the main text with some figures, to show the preferred parameter space and its features. For the scanning parameters given in Table.~\ref{pspace:s} and  Table.~\ref{pspace:f}, the correct relic density can be achieved in a wide parameter space, where the companion-Higgs portal coupling and the DM-companion coupling should be sufficiently large; see the left panels in Fig.~\ref{para:s} and Fig.~\ref{para:f}. The viable DM mass mostly lies above about 200 GeV, and can extend above the TeV region. However, for the upper bound of parameters that we impose, assistant from coannihilation is required, as is explicit in the example of model-B (similar in model-A although not present) shown in the right panel of Fig.~\ref{para:f}: For $M_\Psi\gtrsim 1$TeV, the required degeneracy $\delta\lesssim 4\%$. Therefore, the natural parameter space predicts a sub-TeV DM, if we do not want so large couplings that the theory hits the Landau pole at a quite low scale.

\begin{figure}[htbp]
\centering
\includegraphics[width=0.4\textwidth]{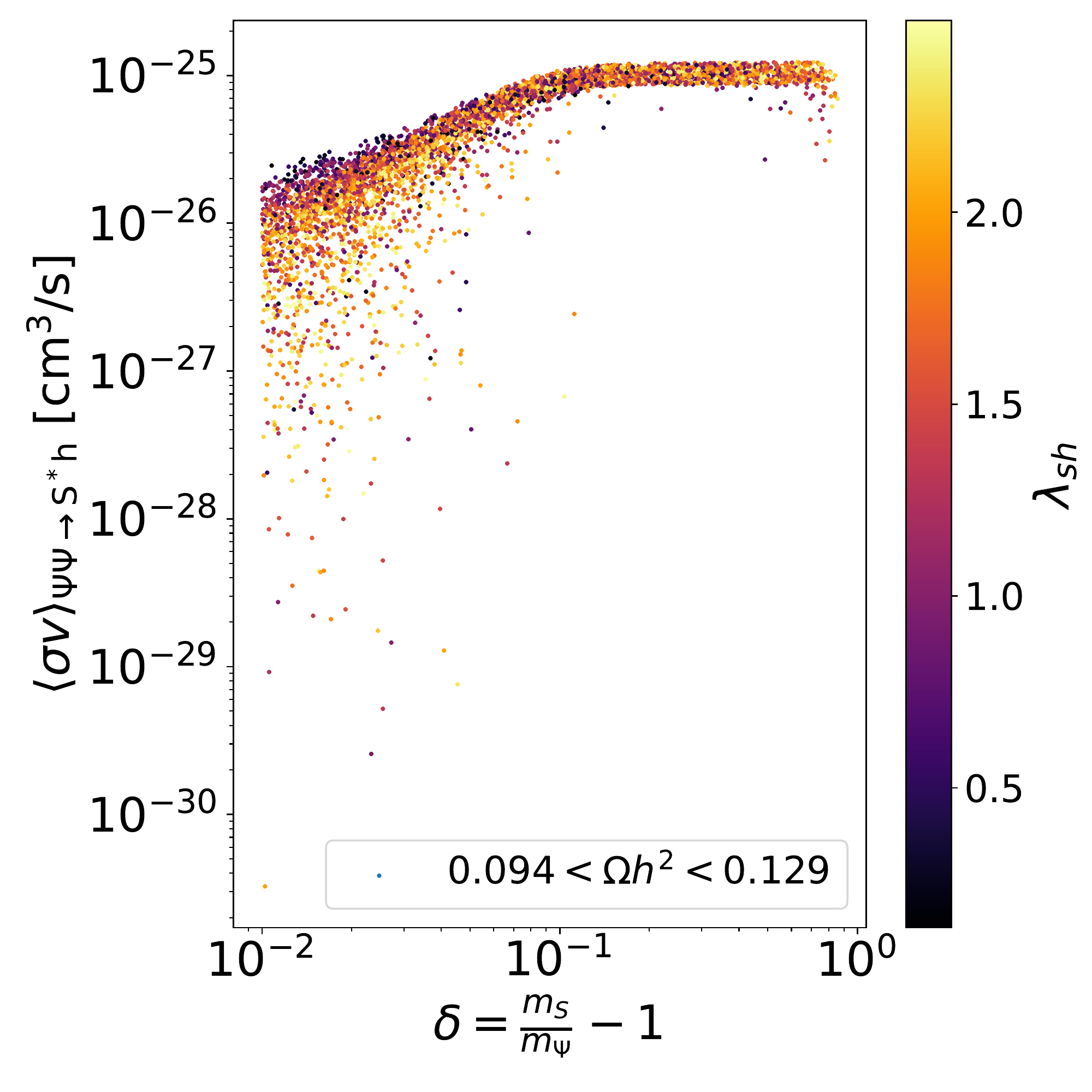}
\includegraphics[width=0.4\textwidth]{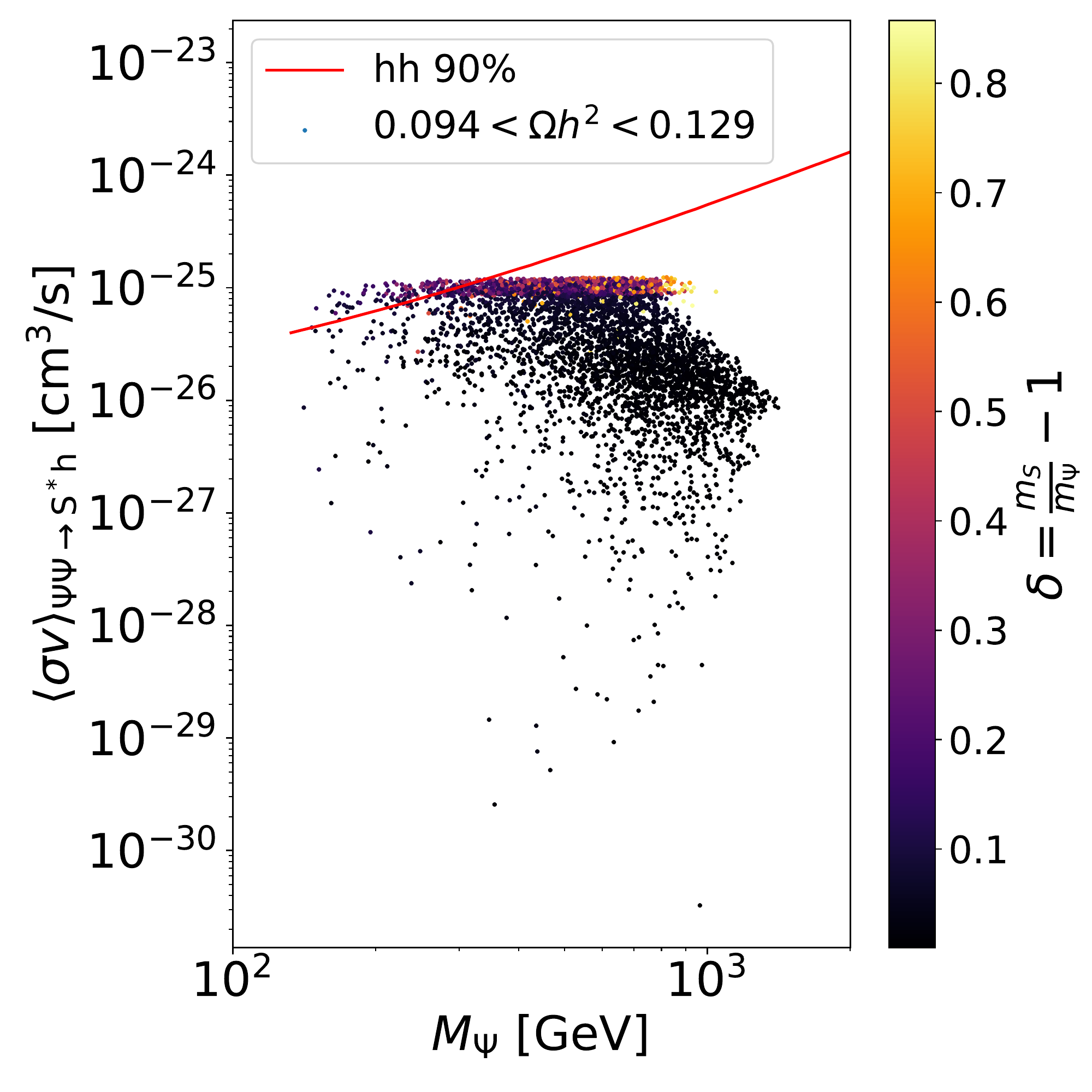}
\caption{Left: In model-B, plot of DM-DM annihilation cross section times DM relative velocity today, depending on the degeneracy $\delta$; the color bar is for the companion-Higgs portal coupling. Right: The schematic constraints on DM via  indirect detection, approximating the DM annihilating spectra as DM+DM$\ra hh$. }
\label{indirect:f}
\end{figure}

To show the indirect detection prospects, in the left panels of Fig.~\ref{indirect:s} and Fig.~\ref{indirect:f} we plot the annihilation cross sections of ${\rm DM +DM}\ra {\rm companion} +h$ time DM velocity today, whose horizontal axis is the degeneracy parameter $\delta$, to highlight the coannihilation effect. As expected, this effect has a significant impact on the indirect detection signals, and from these panels it is seen that $\langle \sigma v\rangle$ may obviously lie below the typical thermal cross section$\sim 1$pb. As explained before, we approximate the spectra of DM annihilation as that of the DM directly annihilating into a pair of Higgs boson. Then, in the left panels of Fig.~\ref{indirect:s} and Fig.~\ref{indirect:f}, we schematically constrain the parameter space. Of interest, the available data is already able to rule out the lighter DM space without coannihilation effect. This situation makes further detailed investigation on the indirect signals of the model necessary, which needs other tools to simulate the spectra.

By the way, in the numerical study on model-A, we did not limit to a very small $\ld_{1h}$, so we show the tight bounds from the DM direct detection in the right panel of  Fig.~\ref{para:s}. Even for $m_{S_1}$ in the TeV region, where the bound is relatively loose, still the allowed $\ld_{1h}\lesssim 0.5$ . As a consequence, the latest PandaX-4T data~\cite{Meng:2021mui} imposing a bound up to 10 TeV (compared to the XENON1T 2018, its exclusion ability is improved by an order of magnitude), rules out the conventional Higgs-portal scenario, which otherwise survived in the high mass region.

\begin{figure}[htbp]
\centering
\includegraphics[width=0.4\textwidth]{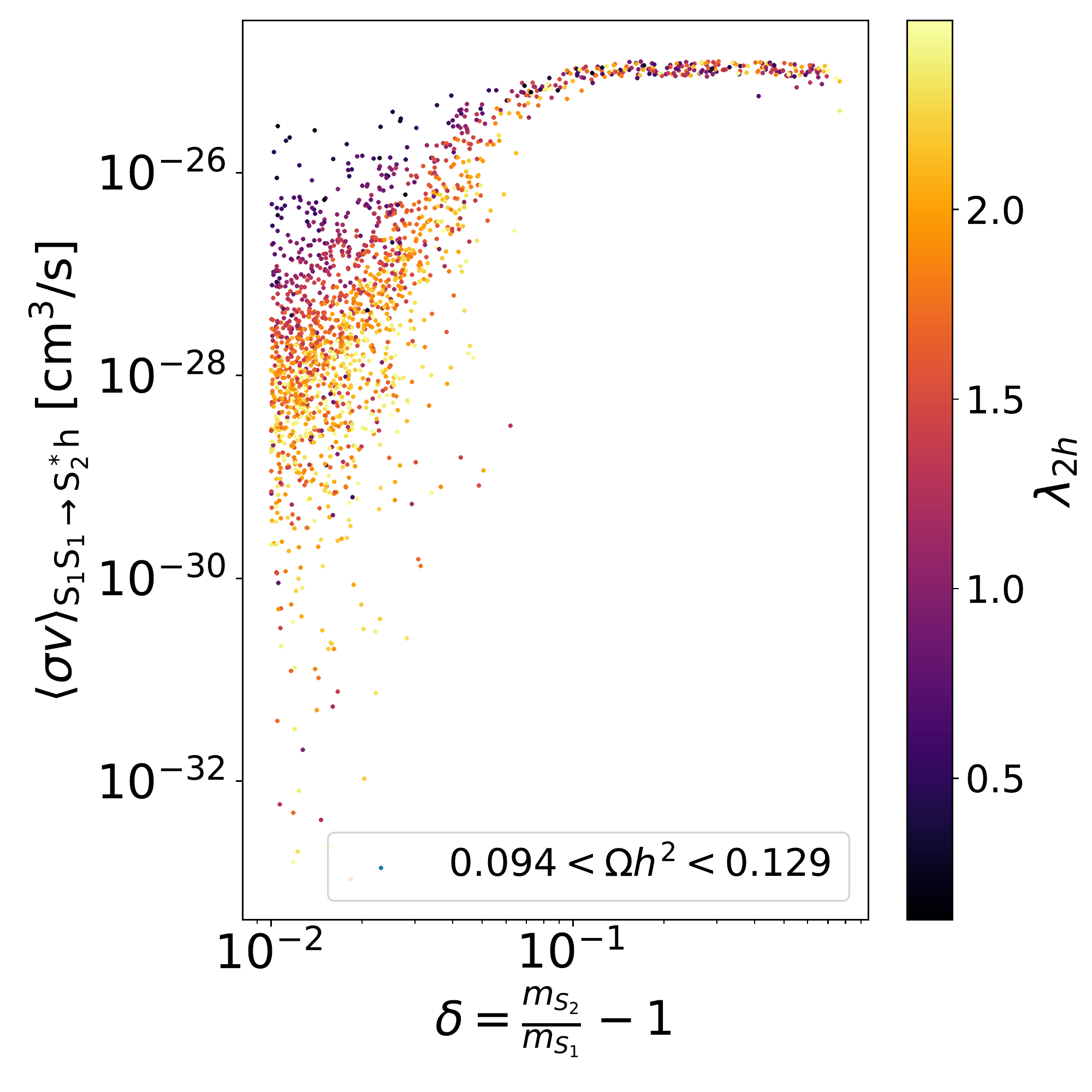}
\includegraphics[width=0.4\textwidth]{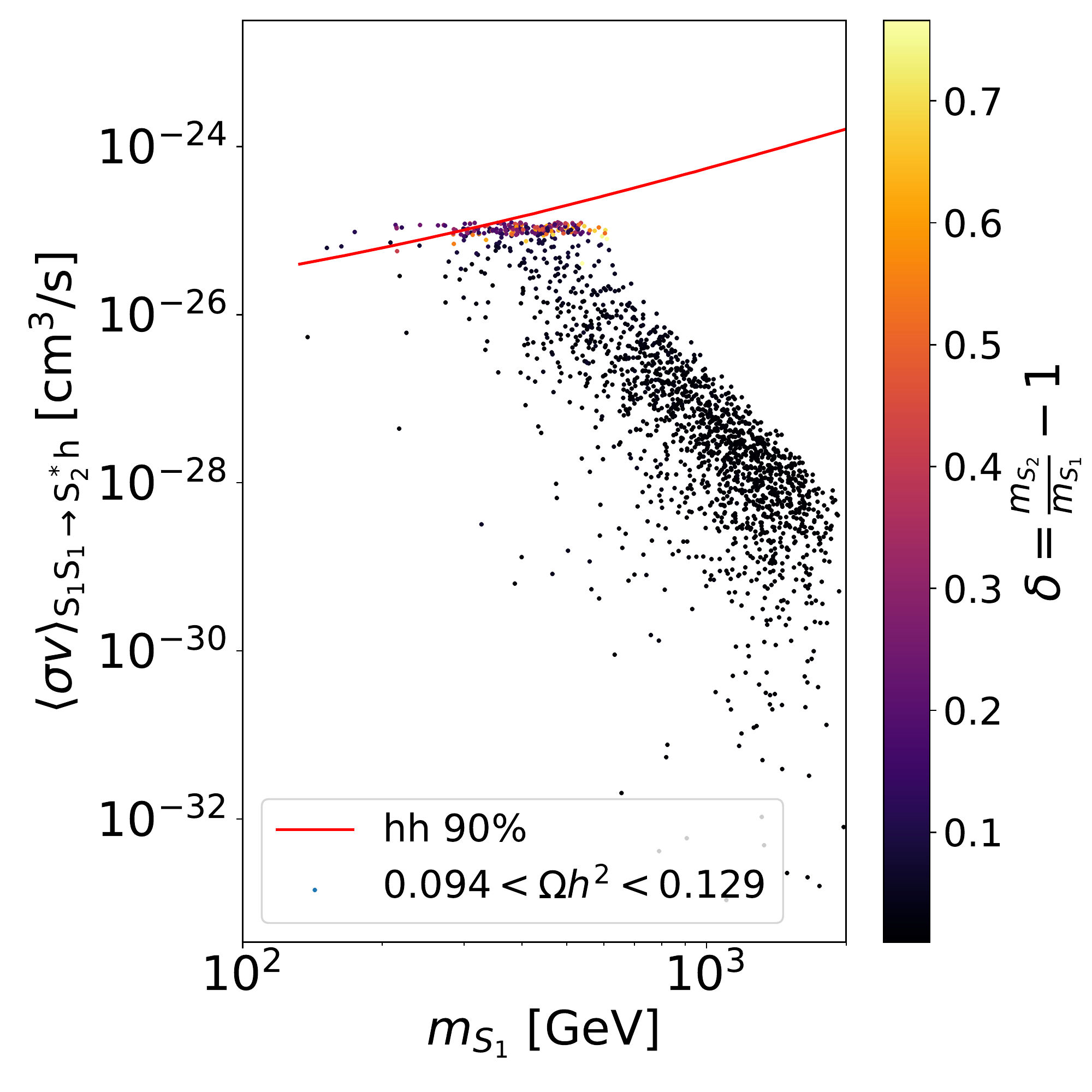}
\caption{Left: In model-A, plot of DM-DM annihilation cross section times DM relative velocity today, depending on the degeneracy $\delta$; the color bar is for the companion-Higgs portal coupling. Right: The schematic constraints on DM via  indirect detection, approximating the DM annihilating spectra as DM+DM$\ra hh$.  }
\label{indirect:s}
\end{figure}

\section{Conclusions and discussions}

Because the correct relic abundance can be obtained naturally, WIMP has been expected to be the leading candidate for dark matter. Nevertheless, facing the more and more stringent constraints from many DM direct detection experiments, the typical WIMP dark matter candidate becomes less and less possible. Hence, in order to save the WIMP DM scenario, we should consider a more complicated structure of the WIMP dark sector.

In this article we propose a novel approach to overcome the strong DM direct detection bound on the WIMP DM, inspired by the semi-annihilating DM models. The crucial assumption is that the DM has a companion which is also charged under the DM protecting symmetry, and moreover the symmetry allows the companion (usually a spin-0 particle) to play the role of bridge between DM and SM, via the trilinear coupling DM-DM-companion. Then, DM semi-annihilates into the companion to reduce its relic density,  having no DM-neulceon scattering signals. Such an idea can be easily realized in the $Z_N$ symmetric models with $N>2$. If DM is a scalar field, one may have to suppress its Higgs portal coupling by hand, whereas the fermionic DM naturally has vanishing coupling to the Higgs. We stress that this mechanism has the characteristics of co-annihilation, and as a matter of fact its effect becomes necessary near or above the TeV region. This means that maybe it is difficult to detect our dark matter directly and indirectly.  

There are some open and pending  issues. First, in the scenario of semi-annihilation, DM is a complex field, consisting of DM and anti-DM. However, DM freeze-out is characterized by DM-DM instead of DM and anit-DM annihilation, which is traced back to DM number violation from the cubic terms. Therefore, the model provides all three elements to generate an asymmetry between DM and anti-DM if CP conservation is abandoned. Such a very interesting possibility deserves a further study elsewhere. Next, the companion may be also a DM component, which brings a remarkable difference to DM phenomenologies. Third, from the side of model building, it is of interest to explore if the idea can be realized in the case that DM protecting symmetry is a non-Abelian group. Last but not least, looking for imprints of our model in the cosmic ray and accelerator such as LHC are also on the agenda. We commented both aspects in the main text, and moreover used a direct di-Higgs boson to approximate the cosmic ray spectra from DM annihilation,. However, a lot of work needs to be done to obtain reliable spectral lines

\appendix 

\section{DM-DM semi-annihilation cross section in model-B}\label{cross:calu}

In this appendix we give the details of the cross section of DM annihilation in model-B with a fermionic DM. The Lagrangian is given by Eq.~(\ref{Z3:fermion}). We list below the relevant interactions and the corresponding Feynman rules:
\begin{align}
\lambda_{L}\bar{\Psi}^{c}P_L\Psi{S}:\lambda_L(1-\gamma5),\\
\lambda_{R}\bar{\Psi}^{c}P_R\Psi{S}:\lambda_R(1+\gamma5),\\
\lambda_{L}^*\bar{\Psi}P_L\Psi^{c}{S}:\lambda_L^*(1+\gamma5),\\
\lambda_{R}^*\bar{\Psi}P_R\Psi^{c}{S}:\lambda_R^*(1-\gamma5).
\end{align}
With them we can calculate the usual spin averaged amplitude squared for DM-DM annihilation:
\begin{align}
\frac{1}{4}{\sum_{spins}|M|^2}&=2\frac{\lambda^2_{sh}(\lambda_L^2+\lambda_R^2-2\lambda_L\lambda_R)v^2m^2_{\chi}}{(q^2-m^2_s)^2},
\end{align}and then the cross section times the DM relative velocity is
\begin{align}
\sigma_0 v &= \frac{|\vec{p}_h|(\lambda_L^2+\lambda_R^2-2\lambda_L\lambda_R)\lambda_{sh}^2v^2m_{\chi}^2}{2\pi\sqrt{s}^3(s-m_s^2)^2},
\end{align}
where $s$ is the square of the energy of the center of mass. Noe that the above expression includes both (putatively equal)  cross sections for DM-DM and ${\rm DM}^*-{\rm DM}^*$ annihilation.

But for the fermonic DM, to obtain the cross section that appears in the formula for calculating relic density, 
\begin{align}
\Omega h^2 = 1.69 \times \frac{x_f}{20}\sqrt{100/g_*} \left(\frac{10^{-10}\rm GeV^{-2}}{\langle\sigma v\rangle} \right)\label{relic}
\end{align}
with $g_*$ the total degree of freedom, we should handle the spin degrees of freedom properly. In the density expression of the initial state, a Dirac fermion has 4 degrees of freedom, including 2 spin states for each of the fermion $\psi$ and an antifermion $\bar{\psi}$. According to the equipartition theorem, we know $n_i=\frac{n}{4}$ for every spin type particle, with $n$ the total particle number density. Moreover, the number density of each spin degree evolves as $\frac{\mathrm{d}n_i}{\mathrm{d}t}=-\sum_{j=1}^4\langle \sigma_{ij}v\rangle n_i n_j$. Then, summing over $i$, we get the following equation~\cite{Srednicki:1988ce},
\begin{align}
   \frac{\mathrm{d}n}{\mathrm{d}t}&=-\frac{1}{16}\sum_{i,j=1}^4\langle \sigma_{ij}v\rangle n^2=-\langle \sigma v \rangle n^2,
\end{align}
where we have defined the cross section
\begin{align}
    \sigma \equiv \frac{1}{16} \sum_{i,j=1}^4 \sigma_{ij}= \frac{1}{16} \sum_{spins}\L \sigma_{\bar{\Psi}\Psi}+\sigma_{{\Psi}\bar\Psi}+\sigma_{ {\Psi}\Psi}+\sigma_{\bar{\Psi}\bar\Psi}\R.
\end{align}
Specific to our model, there are only two types of reactions in our model, namely $\sigma_{\bar{\Psi}\Psi}=\sigma_{\Psi\bar{\Psi}}=0$, and moreover we assumed $\sigma_{ {\Psi}\Psi}=\sigma_{\bar{\Psi}\bar\Psi}$. Therefore, we can establish the relationship between $\sigma$ and $\sigma_0$ as follows:
\begin{align}
  \sigma &=\frac{1}{16}\sum_{spins}(\sigma_{\bar{\Psi}\bar{\Psi}}+\sigma_{\Psi\Psi})=\frac{1}{4}\times\frac{1}{4}\sum_{spins}(\sigma_{\bar{\Psi}\bar{\Psi}}+\sigma_{\Psi\Psi})=\frac{1}{4}\sigma_0.
\end{align}
Eventually, we obtain the cross section adopt in the relic density formula Eq.~(\ref{relic}),
\begin{align}
\sigma v&=\f{1}{4}\sigma_0 v= \frac{|\vec{p}_h|(\lambda_L^2+\lambda_R^2-2\lambda_L\lambda_R)\lambda_{sh}^2v^2m_{\chi}^2}{8\pi\sqrt{s}^3(s-m_s^2)^2}.
\end{align}

\noindent {\bf{Acknowledgements}}

This work is supported in part by the National Science Foundation of China (11775086).


\vspace{-.3cm}
\begin{thebibliography}{99}



\bibitem{Bertone:2004pz}
G.~Bertone, D.~Hooper and J.~Silk,
Phys. Rept. \textbf{405}, 279-390 (2005).

\bibitem{Planck:2018vyg}
N.~Aghanim \textit{et al.} [Planck],
Astron. Astrophys. \textbf{641}, A6 (2020)
[erratum: Astron. Astrophys. \textbf{652}, C4 (2021)].


\bibitem{Aprile:2018dbl}
E.~Aprile \textit{et al.} [XENON],
Phys. Rev. Lett. \textbf{121}, no.11, 111302 (2018). 

\bibitem{Meng:2021mui}
Y.~Meng, Z.~Wang, Y.~Tao, A.~Abdukerim, Z.~Bo, W.~Chen, X.~Chen, Y.~Chen, C.~Cheng and Y.~Cheng, \textit{et al.}
[arXiv:2107.13438 [hep-ex]].

\bibitem{McDonald:2001vt} 
  J.~McDonald,
  Phys.\ Rev.\ Lett.\  {\bf 88}, 091304 (2002)
  [hep-ph/0106249].

  
\bibitem{Hall:2009bx} 
  L.~J.~Hall, K.~Jedamzik, J.~March-Russell and S.~M.~West,
  JHEP {\bf 1003}, 080 (2010).
 
\bibitem{Bernal:2017kxu}
N.~Bernal, M.~Heikinheimo, T.~Tenkanen, K.~Tuominen and V.~Vaskonen,
Int. J. Mod. Phys. A \textbf{32}, no.27, 1730023 (2017).


\bibitem{DEramo:2010keq}
F.~D'Eramo and J.~Thaler,
JHEP \textbf{06}, 109 (2010).

\bibitem{Gross:2017dan}
C.~Gross, O.~Lebedev and T.~Toma,
Phys. Rev. Lett. \textbf{119}, no.19, 191801 (2017).

\bibitem{Okada:2020zxo}
N.~Okada, D.~Raut and Q.~Shafi,
Phys. Rev. D \textbf{103}, no.5, 055024 (2021).

\bibitem{Jiang:2019soj}
X.~M.~Jiang, C.~Cai, Z.~H.~Yu, Y.~P.~Zeng and H.~H.~Zhang,
Phys. Rev. D \textbf{100}, no.7, 075011 (2019). 


\bibitem{Belanger:2012vp}
G.~Belanger, K.~Kannike, A.~Pukhov and M.~Raidal,
JCAP \textbf{04}, 010 (2012). 

\bibitem{Belanger:2014bga}
G.~B\'elanger, K.~Kannike, A.~Pukhov and M.~Raidal,
JCAP \textbf{06}, 021 (2014). 


\bibitem{Ko:2014nha}
P.~Ko and Y.~Tang,
JCAP \textbf{05}, 047 (2014). 


\bibitem{Chen:2017kvz}
S.~L.~Chen and Z.~Kang,
JCAP \textbf{05}, 036 (2018). 

\bibitem{Kang:2014cia}
Z.~Kang,
Eur. Phys. J. C \textbf{75}, no.10, 471 (2015). 


\bibitem{Feng:2011vu}
J.~L.~Feng, J.~Kumar, D.~Marfatia and D.~Sanford,
Phys. Lett. B \textbf{703}, 124-127 (2011).  

\bibitem{Kang:2010mh}
Z.~Kang, T.~Li, T.~Liu, C.~Tong and J.~M.~Yang,
JCAP \textbf{01}, 028 (2011). 

\bibitem{Gao:2011ka}
X.~Gao, Z.~Kang and T.~Li,
JCAP \textbf{01}, 021 (2013).

\bibitem{Saez:2021oxl}
B.~D\'\i{}az S\'aez, K.~M\"ohling and D.~St\"ockinger,
[arXiv:2103.17064 [hep-ph]].

\bibitem{Belanger:2011ww}
G.~Belanger and J.~C.~Park,
JCAP \textbf{03}, 038 (2012).  

\bibitem{Yaguna:2019cvp}
C.~E.~Yaguna and \'O.~Zapata,
JHEP \textbf{03}, 109 (2020). 



\bibitem{Ma:2007gq}
E.~Ma,
Phys. Lett. B \textbf{662}, 49-52 (2008). 

\bibitem{Aoki:2014cja}
M.~Aoki and T.~Toma,
JCAP \textbf{09}, 016 (2014). 


\bibitem{Guo:2015lxa}
J.~Guo, Z.~Kang, P.~Ko and Y.~Orikasa,
Phys. Rev. D \textbf{91}, no.11, 115017 (2015). 


\bibitem{Griest:1990kh}
K.~Griest and D.~Seckel,
Phys. Rev. D \textbf{43}, 3191-3203 (1991)
doi:10.1103/PhysRevD.43.3191



\bibitem{Kang:2017mkl}
Z.~Kang, P.~Ko and T.~Matsui,
JHEP \textbf{02}, 115 (2018). 

\bibitem{Kannike:2019mzk}
K.~Kannike, K.~Loos and M.~Raidal,
Phys. Rev. D \textbf{101}, no.3, 035001 (2020). 

\bibitem{Chiang:2019oms}
C.~W.~Chiang and B.~Q.~Lu,
JHEP \textbf{07}, 082 (2020). 

\bibitem{Jungman:1995df}
G.~Jungman, M.~Kamionkowski and K.~Griest,
Phys. Rept. \textbf{267}, 195-373 (1996). 

\bibitem{mickay}
David J. C. MacKay,
Information Theory, Inference, and Learning Algorithms



\bibitem{Belanger:2020gnr}
G.~Belanger, A.~Mjallal and A.~Pukhov,
Eur. Phys. J. C \textbf{81}, no.3, 239 (2021)


\bibitem{gap method}
S.Yellin,
Phys. Rev. D66, 032005 (2002).


\bibitem{Abazajian:2020tww}
K.~N.~Abazajian, S.~Horiuchi, M.~Kaplinghat, R.~E.~Keeley and O.~Macias,
Phys. Rev. D \textbf{102} (2020) no.4, 043012


\bibitem{Martin:1997ns}
S.~P.~Martin,
Adv. Ser. Direct. High Energy Phys. \textbf{18}, 1-98 (1998).

\bibitem{Srednicki:1988ce}
M.~Srednicki, R.~Watkins and K.~A.~Olive,
Nucl. Phys. B \textbf{310}, 693 (1988).



\end{thebibliography}
\end{document}